\documentclass[onecolumn,twoside]{article}
\makeatletter

\input epsf.tex

\begin{document}
\title{SEMI-AUTOMATED S\`{E}NARMONT METHOD FOR MEASUREMENT OF SMALL RETARDATION}
\author{Atsushi Mori$^1$\thanks{Corresponding author. E-mail: atsushimori@tokushima-u.ac.jp\vspace{6pt}},
Ryosuke Tomita$^2$}
\date{
\begin{center}
\begin{small}
$^1${}\textit{Department of Advanced Materials, Institute of Technology and Science, Tokushima University, Tokushima 770-8506, Japan} \\
$^2${}\textit{Department of Optical Systems Engineering, Collage of Systems Innovation Engineering, Tokushima University, Tokushima 770-8506, Japan} \\[1ex]
(\textit{Received July 5, 2014; revised December 18, 2014; in final form\hspace{3cm}})
\end{small}
\end{center}}
\maketitle
\begin{small}
In usual measurement with the S\`{e}narmont method using a conventional polarization microscope, the azimuth angle of the analyzer at extinction of the emerging light is detected by naked eyes.
If the intensity of light is measured as a function of the azimuth angle of the analyzer, one can find the direction at which the intensity is minimized more accurately by fitting.
However, this procedure requires much numbers of operations as compared to the usual method, and thus is time-consuming.
To circumvent this problem a setting with which the intensity is measured as a function of the azimuth angle of the analyzer is constructed with rotation of the analyzer through a computer control.
As a result, the time required for the measurement has been greatly decreased. \\[2ex]
\end{small}
%
%
\newpage
\section*{INTRODUCTION
\label{sec:intro}}
Birefringence is a measure of structural anisotropy of materials.
For example, the measure of birefringence is proportional to the nematic order parameter in nematic liquid crystals.$^{\mbox{\footnotesize\cite{Strobl}}}$
Form and flow birefringence reflects the structural anisotropy of polymers.$^{\mbox{\footnotesize\cite{Doi}}}$
Stress of transparent materials can be evaluated through the measurement of stress birefringence.$^{\mbox{\footnotesize\cite{Born,Hect}}}$

Equipment of the S\`{e}narmont method is usually used with a polarization microscope.
A linear polarizer is located after the light source to make a linearly polarized light, which incidents into a sample.
After transmittance through the sample an elliptically polarized light incidents into a quarter wave plate.
This elliptically polarized light is transformed into a linearly polarized one by the quarter wave plate.
If the trajectory of electric field of the elliptically polarized light is $E_x^2/A_x^2 + E_y^2/A_y^2 = 1$, the direction of vibration of the electric field of the linearly polarized light emerging from the quarter wave plate, whose fast axis coincides to the transmission direction of the polarizer, is rotated by the angle $\theta \equiv \tan^{-1} A_y/A_x$.
One rotates the analyzer to seek the angle of the transmission direction of the analyzer for which the emerging light from the analyzer is extinguished.
Hereafter, the rotation angle of transmission direction is termed as the azimuth angle.

Although the S\`{e}narmont method$^{\mbox{\footnotesize\cite{Senarmont1840}}}$ is old method, improvement of this method was sometimes revisited even in recent years.
Usually determination of the phase difference $\delta$ accompanies uncertainty of $2\pi m$ with $m$ being an integer.
Measurement of a retarder with $|\delta|$ larger than $2\pi$ was a subject of some recent studies$^{\mbox{\footnotesize\cite{Kaufmann2002,Kaufmann2003,Nagib2003}}}$.
A multicolor S\`{e}narmont method, in which one could determine the order of the retardation by measurements alone, was employed in Refs.~\cite{Kaufmann2002,Kaufmann2003}.
On the other hand, the wavelength dependence of the retardation based on the order of the retardation at a standard wavelength was estimated in Ref.~\cite{Nagib2003}.
As opposed to the large retardation, the present authors concentrate on a specimen with a tiny retardation.
Limiting to the small retardation, this uncertainty can be avoided.
Automation of the measurement was a subject of Refs.~\cite{Kaufmann2003,Nagib2003,Ajovalasit2011}, which is also the main aim of the present study.
In these papers as well as in the present study, it is unnecessary to seek the azimuth angle of the analyzer at extinction.
A fast measurement was a subject of Ref.~\cite{Kaufmann2003}, which is accomplished in the present study. 
Formulation on usage of an incorrect quarter wave plate in the S\`{e}narmont method, which will be reproduced in Jones formalism in this paper, was given already.$^{\mbox{\footnotesize \cite{Ratajczyk2000,Kurzynowski2001,Ratajczyk2002}}}$

Sensitivity of birefringence measurement in the S\`{e}narmont method was already improved by fitting of the intensity of light emerging from the analyzer by one of the present authors and co-workers.$^{\mbox{\footnotesize\cite{Mori2008}}}$
The measure of birefringence $\Delta n$ as small as $10^{-8}$ were successfully measured for $d$ $\sim$ 10 mm thick sample of silica gels prepared in a high magnetic field of various strengths such as $B$ $\le$ 5 T.
In other words, retardations of order of $|\mit\Gamma|$ ($=$ $|\Delta n d|$) $\sim$ 10 nm was successfully measured.
This value can be identified to the order of magnitude of sensitivity; for accurate determination one should measure standard samples with known retardation.
The intensities $I$ were measured at every $10^\circ$ of the azimuth angle of the analyzer $\theta$.
Measurements at nineteen $\theta$'s was done and the intensity $I(\theta)$ was fitted by
\begin{equation}
\label{eq:fitMori}
I(\theta) = A_1 \cos(2\theta-\delta) + A_2,
\end{equation}
and the retardation $\mit\Gamma$ $\equiv$ $\Delta n d$ was calculated through the phase difference due to the birefringence $\delta$ $=$ $2\pi \mit\Gamma/\lambda$.
A mercury lamp is used as a light source commonly with a polarization microscope.
In a previous study,$^{\mbox{\footnotesize\cite{Mori2008}}}$ also a mercury light source was used.
So, the wavelength of the light is $\lambda$ $=$ 546 nm.
The intensity $I$ at an azimuth angle was measured by a spectrometer spending a few minutes.
Thus, it took an hour or so for measurement of $I(\theta)$ at one point in a sample.

It should be emphasized that the cost is another factor in practice, such as mentioned in Ref.~\cite{Velasquez2005}.
A materials researcher, or a researcher in fields other than the optics, unnecessarily has a set of equipment, which enables a fairly fast measurement.
Lowering the cost for the equipment is another important factor.
Even if the cost unreasonably high, one needs a time to get budget and the study does not go well.
This issue has been resolved in the preset work, too.

This paper reports that even in the classical optical method one can achieve a highly sensitive measurement of birefringence.
The sensitivity and potentially the accuracy can be improved a few order of magnitude.
This achievement does not rely on the new physical principle.
In other words, instrumentation itself has been shown to have a potential to develop a measurement technique.
Again, it is stressed that a highly sensitive measurement of birefringence is possible with a low cost equipment.

\section*{PRINCIPLE
\label{sec:principle}}
Configuration of the S\`{e}narmont method is as shown in Figure~\ref{fig:Senarmont}.
$z$-axis is taken along the ray.
The transmission direction of the polarizer is set along $x$-axis.
The sample is mounted after the polarizer at $45^\circ$ from $x$-axis.
A quarter wave plate is inserted between the sample and the analyzer with its fast axis along $x$-axis.
The analyzer is aligned at angle $\theta$ with respect to $y$-axis.

The intensity is calculated here as a function of the azimuth angle of the analyzer $\theta$.
The Jones vector$^{\mbox{\footnotesize\cite{Jones}}}$ of the light emerging from the analyzer is given by
\begin{eqnarray}
\nonumber
J &=&
\left[
\begin{array}{cc}
\cos\theta & -\sin\theta \\
\sin\theta & \cos\theta
\end{array}
\right]
\left[
\begin{array}{cc}
0 & 0 \\
0 & 1
\end{array}
\right]
\left[
\begin{array}{cc}
\cos\theta & \sin\theta \\
-\sin\theta & \cos\theta
\end{array}
\right]
\left[
\begin{array}{cc}
1 & 0 \\
0 & e^{i\gamma}
\end{array}
\right] \\
&\times&
\frac{1}{2}
\left[
\begin{array}{cc}
1 & 1 \\
-1 & 1
\end{array}
\right]
\left[
\begin{array}{cc}
e^{-i\delta/2} & 0 \\
0 & e^{i\delta/2}
\end{array}
\right]
\left[
\begin{array}{cc}
1 & -1 \\
1 & 1
\end{array}
\right]
\left[
\begin{array}{c}
1 \\
0
\end{array}
\right].
\end{eqnarray}
Here, the Jones matrix of the sample, which is regarded as a retarder of the phase difference $\delta$ with the fast axis aligned at $45^\circ$, is
\begin{displaymath}
\frac{1}{2}
\left[
\begin{array}{cc}
1 & 1 \\
-1 & 1
\end{array}
\right]
\left[
\begin{array}{cc}
e^{-i\delta/2} & 0 \\
0 & e^{i\delta/2}
\end{array}
\right]
\left[
\begin{array}{cc}
1 & -1 \\
1 & 1
\end{array}
\right],
\end{displaymath}
where
$\frac{1}{\sqrt{2}}\left[
\begin{array}{cc}
1 & \mp 1 \\
\pm 1 & 1
\end{array}
\right]$
is the matrix of rotation of $\mp 45^\circ$.
The Jones matrix of the quarter wave plate is expressed as
$\left[
\begin{array}{cc}
1 & 0 \\
0 & e^{i\gamma}
\end{array}
\right]$
with $\gamma$ begin $90^\circ$ for an ideal quarter wave plate.
In actual, the quarter wave plate is not ideal.
So, $\gamma$ does not perfectly coincide to $90^\circ$.
Evaluation of $\gamma$ taking into account the dispersion will be given in ^^ ^^ Quarter Wave Plate" subsection.
The intensity is calculated by $I$ $\propto$ $^tJ^{*} \cdot J$ as
\begin{equation}
I(\theta) \propto \frac{1}{2} - \frac{1}{2} (\cos2\theta\cos\delta +\sin\gamma\sin 2\theta\sin\delta). 
\end{equation}
By introducing the background intensity and the intensity of the light source as fitting parameters, the following fitting equation is obtained.
\begin{equation}
\label{eq:fitnew}
I(\theta) = A_1(\cos2\theta\cos\delta +\sin\gamma\sin 2\theta\sin\delta) + A_2.
\end{equation}
It is noted that this equation is reduced to Equation (\ref{eq:fitMori}) at $\gamma$ $=$ $90^\circ$.
By fitting the measured intensities $I$'s at several $\theta$'s by Equation (\ref{eq:fitnew}) with $A_1$, $A_2$, and $\delta$ as fitting parameters, the phase difference due to birefringence $\delta$ is obtained.
One can calculate the measure of birefringence $\Delta n$ through the retardation $\mit\Gamma$ via
\begin{equation}
\label{eq:retardation}
\delta \mbox{[rad]} =\frac{2\pi \mit\Gamma}{\lambda}.
\end{equation}

\section*{RESULTS
\label{sec:results}}
\subsection*{Setup
\label{sec:conf}}
Traversing along optical axis ($z$-axis), a light source, the polarizer (P), the sample folder (S), the quarter wave plate (Q), a diaphragm (D), and the analyzer (A) with a mechanical rotator are arranged as shown in Figure~\ref{fig:Senarmont}.
A light source, which was a He-Ne laser in this study, is located before P, and a photo detector is located beyond A.
A specimen is mounted on S with the long axis of the cell vertical.
It means that if the optic axis of the sample --- in a case that the sample prepared in an external field the direction of the field becomes the optic axis from a symmetrical consideration ---  is vertical, then $x$-axis is along $45^\circ$ from right horizontal direction counter clockwise viewing along the optical axis of the system from the detector.
If the optic axis of the sample is horizontal, then $x$-axis is set along $45^\circ$ from the right horizontal direction clockwise.
It is note here that in a previous work$^{\mbox{\footnotesize\cite{Mori2008}}}$ the magnetic field was applied along the long axis of the sample cell.
Contrary, for compare of the difference between properties along the direction of the magnetic field and those across that the samples were prepared with the magnetic field applied normal to the long axis of the cell.$^{\mbox{\footnotesize\cite{Mori2014}}}$
Anyway, the transmission direction of P is set along $x$-axis.
Also the fast axis of Q is set along $x$-axis.
$y$-axis is defined so that the three axes make a right-handed system.
The components P, S, Q, A, and the rotator are purchased from Sigma Koki Co., Ltd. (the polarizers for P and A are SPF-30C-32, the wave plate for Q is WPM-20-4P, the folders for P and Q are respectively PH-20 and SPH-30, the rotation stage for A is SGSP-40YAW).  
The nominal accuracy of the rotation stage of A is $0.005^\circ$ per pitch.

The output of the photo detector is transferred to a module which makes analog-digital (AD) conversion and then put this result to a USB port of the computer.
This module was composed of a USB serial converter (FT323RL, FTDI Ltd.), a 10-bit 2-channel AD converter (MCP3002, Microchip Technology Inc.), and a CMOS/RS232C level converter (ADM3202, Analog Devices, Inc.).
A stepping motor controller (PAT-001, Sigma Koki Co., Ltd.), which controls the rotator, is also connected to it in order to obtain the azimuth angle of A.
The intensity $I$ and the azimuth angle $\theta$ are simultaneously put into the computer. 
In this way, the intensity of the light impinging on the detector is measured as a function of $\theta$.
The photo detector is composed of a photo diode (PD) and an operational amplifier (OP AMP) with a register.
The PD and the OP AMP were products of Hamamatsu Photonics K. K. (S2386-45K) and National Semiconductor Co. (LM741CN), respectively.
Although date of the photosensitivity of PD at room temperature is available from the web site,$^{\mbox{\footnotesize\cite{Hamamatsu}}}$ data for linearity does not appear on this url (the light intensity at which the photosensitivity curve was measured is not specified).
Data of the OP AMP is also available,$^{\mbox{\footnotesize\cite{TI}}}$ but the situation is the same.
Below, discussion will be given assuming the linearities of PD and OP AMP are enough.
In other words, fluctuation of intensity of light source is not a literal one but a comprehensive one. 
The OP AMP was driven by a $\pm15$ V power source.
The driver of the hardware and the header of the program of controlling the system are obtained from Future Technology Devices International (FTDI) Ltd.$^{\mbox{\footnotesize\cite{FTDI}}}$

\subsection*{Quarter Wave Plate
\label{sec:QWP}}
The phase differences of the quarter wave plate Q at various wavelengths, which are obtained from Sigma Koki Co., Ltd.$^{\mbox{\footnotesize\cite{Sigma}}}$, is plotted in Figure~\ref{fig:QWP} by dots.
Those data are well fitted by the Cauchy's equation $\gamma \mbox{[$^\circ$]}$ $=$ $a/(\lambda \mbox{[nm]})$ $+$ $b/(\lambda \mbox{[nm]})^3$ with $a$ $=$ 52232, $b$ $=$ $-2.535\times10^7$, as shown in the figure by a curve.
From this fitting result, an interpolated values $\gamma$ $=$ $82.4^\circ$ ($\sin\gamma$ $=$ 0.991) is obtained at the working wave length $\lambda$ $=$ 632.8 nm.
The same result could be obtained by fitting by a third order polynomial of $\lambda$ or $\gamma \mbox{[$^\circ$]}$ $=$ $a$ $+$ $b/\lambda^2$ $+$ $c/\lambda^3$.

\subsection*{Measurements
\label{sec:Measurements}}
After making measurement without sample to evaluate the zero point error, a measurement of a sheet of Scotch tape have been examined.
It has taken only a few minutes to measure at one point.
The time required for measurement have been greatly decreased as compared with a previous method$^{\mbox{\footnotesize\cite{Mori2008,Mori2014}}}$.

One of the measurement data without sample were plotted in Figure~\ref{fig:zero}.
Because a calibration have not been made, the vertical axis is of arbitrary unit.
To obtain one intensity data, after measurements for determining the initial azimuth angle 50 measurements were performed.
The standard deviations over 50 measurements were drawn as the full error bars.
Those errors originated mainly in the fluctuation of intensity of light source (the fluctuation of intensity of a laser is significantly smaller than that of, for example, a mercury light), but affected by the properties of PD, OP AMP, AD converter and so on, as mentioned in ^^ ^^ Setup" subsection. 
The result of fitting by Equation (\ref{eq:fitnew}) is $A_1$ $=$ $-189$, $A_2$ $=$ 340, $\delta$ $=$ $-0.007^\circ$.
For zero point error estimation in this work, as one set of measurements, ten measurements for initial azimuth angle determination and then ten for $I(\theta)$ were done; the initial azimuth angle $\theta_0$ was determined by $\tan2\theta_0$ $=$ $\sin\gamma\tan\delta$ (Equation (\ref{eq:fitnew}) is minimized when $\tan2\theta$ $=$ $\sin\gamma\tan\delta$) for the former ten measurements and then the averaged $\theta_0$ was subtracted from the results of $\theta$ of the latter ten ones prior to the fitting.
Five set of measurements without sample were done.
The average over those five data was $\delta$ $=$ $0.005^\circ$ with the standard deviation $0.014^\circ$.
The maximum deviation in the five measurements was $0.04^\circ$, the half of which is regarded as the zero point error ($\delta$[$^\circ$] $=$ $0.005\pm0.02$).
Relying on Equation (\ref{eq:retardation}), one can rewrite $\delta$ into the retardation.
The present result is
\begin{equation}
\mit\Gamma\mbox{[nm]} = \frac{\delta \mbox{[$^\circ$]}}{360} \lambda = 0.009\pm0.074.
\end{equation}
The zero point error can be estimated as 0.08 nm.

Essentially the same measurements have been done for a sheet of Scotch tape.
Determination of the initial azimuth angle was simplified (a couple of mechanical adjustments was made) and only one measurement was done for $I(\theta)$.
From the fitting result, the phase difference is $\delta$ $=$ $192.78^\circ$, which can be regarded as the phase difference of one sheet of Scotch tape.
This phase difference is rewritten into the retardation on the basis of Equation (\ref{eq:retardation}) as $\mit\Gamma$ $=$ 338.87 nm.
Birefringence measurement of stacked Scotch tapes has been done recently in Ref.~\cite{Belendez2010}.
The measure of birefringence at $\lambda$ $=$ 632.8 nm was evaluated as $\Delta n$ $=$ 0.007650.
The corresponding retardation is $\mit\Gamma$ $=$ 326.4 nm (the thickness $d$ $=$ $42.5 \pm 1.3$ $\mu$m).
Because a Scotch tape is not a standard sample, the properties depend on the lot.
Nevertheless, the present results agrees with this value.
The difference between those values may be regarded as the sum of the fluctuation for one sheet of Scotch tape and the system error.
Here, another set of measurements for the Scotch tape is presented.
Measurements for two, three, and four stacked sheets gave averages $\delta_2$ $=$ $386.57^\circ$, $\delta_3$ $=$ $508.28^\circ$, and $\delta_4$ $=$ $740.03^\circ$, respectively.
The average and the standard deviation over $\delta_n$ for $n$ = 1 to 4 ($\delta_1$ $=$ $192.78^\circ$) are $193.25^\circ$ and $0.33^\circ$, respectively.
Corresponding retardation is $\mit\Gamma\mbox{[nm]}$ $=$ $339.69\pm0.29$.
It indicates that the zero point error of 0.08 nm does not affect the measurement results of large retardation.
Further analyses such as based on fitting $\delta_n$ are possible.
However, the authors wish to postpone the final conclusion to the birefringence of the Scotch tape because the measurements for the Scotch tape were mere a demonstration in this paper.

In the present measurement, the intensities $I$'s at every $\theta$ are automatically obtained.
Fitting by Equation (\ref{eq:fitnew}) and calculation of the retardation on the basis of Equation (\ref{eq:retardation}) are not automated.
In this respect, the authors term ^^ ^^ semi"-automated.

\section*{DISCUSSION
\label{sec:discussion}}
\subsection*{Sensitivity and Accuracy
\label{sec:accuracy}}
The zero point error in the retardation evaluated in the preceding subsection was 0.08 nm for the present setup.
In previous studies,$^{\mbox{\footnotesize\cite{Mori2008,Mori2014}}}$ retardations as small as 10 nm were successfully measured.
The former value is much smaller than the latter one.
However, this value is just a lower bound of the sensitivity.
For accurate determination one should measure standard samples with a known retardation.
Nevertheless, we can say that improvement of the sensitivity is expected compared with the previous one$^{\mbox{\footnotesize\cite{Mori2008}}}$ because the previous operation includes errors made by operation of rotating the analyzer by hands.

Discussion about possible sources of the inaccuracy except for electrical ones is given here.
The angle of the quarter plate has been adjusted so that its fast axis coincides to the transmission direction of the polarizer by hands.
For crossed configuration of the polarizer and the analyzer, the emerging light becomes maximum when the fast (or slow) axis of the quarter plate coincide to the transmission direction of the polarizer.
Even if one measures the intensity of the emerging light, the accuracy of azimuth angle of the quarter wave plate is limited as long as one adjusts the azimuth by hands.
It is known that the error in the direction of the specimen only weakens the intensity of the emerging light without affecting the $\theta$ dependence.
Use of a rotator with computer control for the quarter wave plate can be proposed.
If the intensity is measured as a function of the azimuth angle of the quarter wave plate when no specimen is set, the azimuth of the maximum can be found numerically with the accuracy to the nominal value for the rotator.
This method may improve the accuracy as well as the sensitivity.

\subsection*{Cost
\label{sec:cost}}
The cost for the present measurement system has been 300 thousands JPY (about 3 thousands USD) or so.
This is extremely lower than that for commercially available equipment.
For example, that from Otsuka Electronics Co., Ltd (RE-100) is, at least, 5 million JPY (about 50 thousands USD).

\subsection*{Incorrect Quarter Wave Plate
\label{sec:IQWP}}
In this study, a value of $\gamma$ determined by an interpolation was used.
However, one can treat $\gamma$ as a fitting parameter.
In other words, the present formulation is applicable to the case of an incorrect quarter plate with an unknown phase difference.$^{\mbox{\footnotesize\cite{Nagib2000,Kurzynowski2002}}}$
A criticize was given by Kurzynowski~\textit{et~al.}$^{\mbox{\footnotesize\cite{Kurzynowski2002}}}$ on Nagib's method$^{\mbox{\footnotesize\cite{Nagib2000}}}$ as one should rotate both the polarizer and the analyzer simultaneously to find the extinction of light.
In Kurzynowski \textit{et~al.}'s method two successive rotation, one of which is for a normal S\`{e}narmont method and the other for a reverse S\`{e}narmont method --- the specimen and the quarter wave plate are swapped and the polarizer is rotated ---, is required.
In other words, while in Nagib's method the tow-dimensional space is scanned to seek the extinction, in Kurzynowski \textit{et al.}'s method one needs to sweep one-dimensionally.
However, to measure the intensity at several azimuth directions and to make a fitting, which has been employed presently, is overwhelmingly simple.
A less time-consuming method for unknown $\gamma$ can be proposed based on Equation (\ref{eq:fitnew}).

\section*{CONCLUDING REMARKS
\label{sec:conclusion}}
The results of this study are summarized as follows.
\begin{itemize}
\item A setting of a semi-automated S\`{e}naront method has been made.
As a result, the time required for the measurement has been greatly reduced as compared to a previous method$^{\mbox{\footnotesize\cite{Mori2008}}}$.
\item Discussion on the sources of inaccuracy has been given and a method to improve one of them has been proposed.
\item A method for the case of the incorrect quarter wave plate with an unknown phase difference has been proposed.
\item Lowering of the cost has been accomplished.
\end{itemize}

The authors wish to leave the study of the method to reduce the errors, using a computer-controlled rotator for the quarter wave plate, as a future research.
Also, the authors wish to postpone the accurate determination of the sensitivity and the accuracy by measuring standard samples with a known retardation.
Retarders with retardation of a few tens of nanometers are expensive such as 200 thousands JPY (about 2 thousands USD), which is comparable to the total cost of the present setup.

In the present study, a fitting has been made for nineteen data.
By increasing the number of data, one may increase the accuracy.
Of course, even if the time necessary for the measurement increases, the cost remains the same.
Alternatively, the minimum requisite to determine unknowns is to have the same number of measured data as the unknowns, such as done in Refs.~\cite{Ajovalasit2011,Wang2008}.
The rotation of the analyzer is, however, not time-consuming with use of a mechanical rotator with computer control.
Thus, the advantage of reducing the number of measured angles for shortening the time for measurement is small.

A mechanical rotator has been employed to rotate the analyzer in this study.
An alternative is a liquid crystal polarization rotator$^{\mbox{\footnotesize\cite{Wang2008}}}$.
Use of an electrical rotation through the application of a voltage to the liquid crystal rotator may be less time-consuming than the use of the mechanical rotator.
While the accuracy of the rotation angle has been guaranteed such as to the product company's nominal one, the applied voltage may fluctuate, resulting in an inaccuracy of the rotation angle due to the liquid crystal rotator.
Considering comprehensively, the mechanical one has been selected.

\section*{ACKNOWLEDGMENT}
The authors thank Prof. N. Goto for providing some electric parts and optical components for constructing the setup.

\newpage
\section*{Figure captions}
\begin{figure}[h]
\caption{\label{fig:Senarmont}
Configuration of the S\`{e}narmont method for measurement of birefringence.}
\end{figure}

\begin{figure}[h]
\caption{\label{fig:QWP}
Retardation v.s. wavelength for the quarter wave plate presently used.}
\end{figure}

\begin{figure}[h]
\caption{\label{fig:zero}
One of data of measurements without sample and the fitting curve by Equation (\ref{eq:fitnew}).}
\end{figure}

\newpage
\centerline{
\epsfxsize=0.8\textwidth
\epsfbox{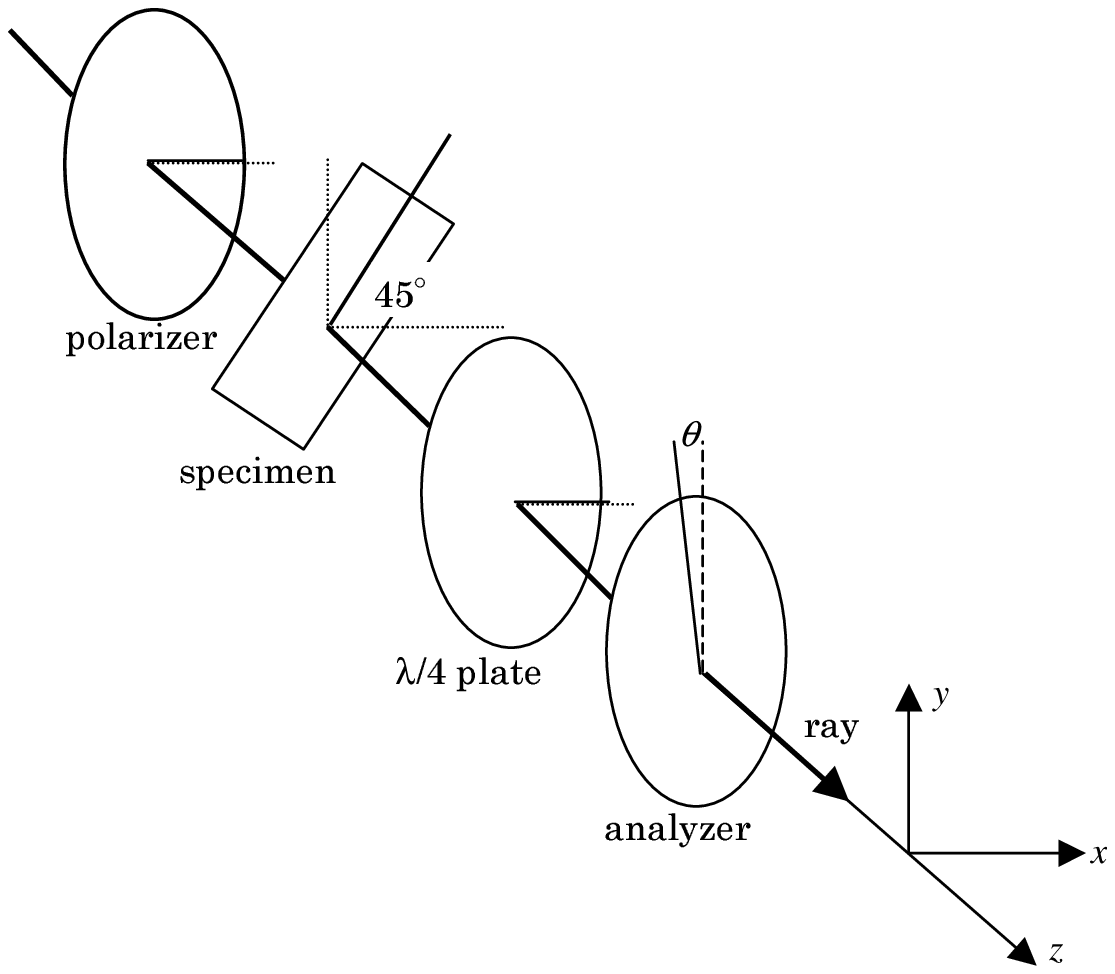}}
\vspace{1cm}
Figure 1

\newpage
\centerline{
\epsfxsize=0.9\textwidth
\epsfbox{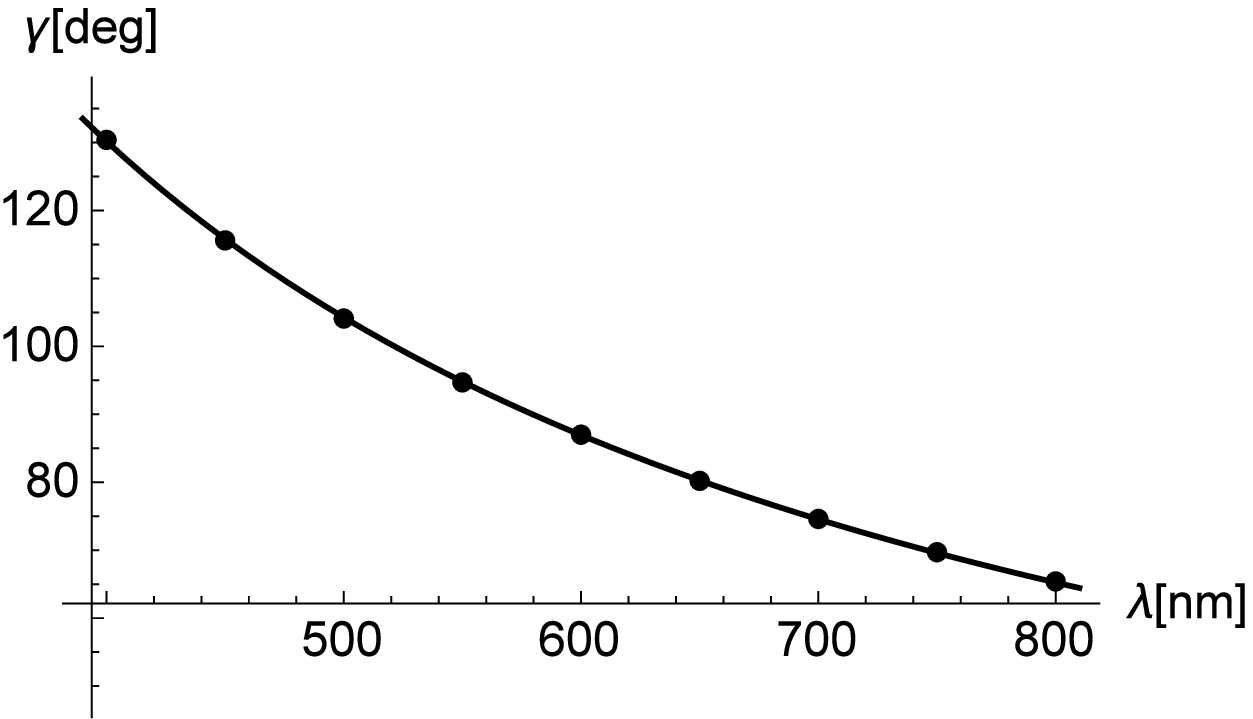}}
\vspace{1cm}
Figure 2

\newpage
\centerline{
\epsfxsize=0.9\textwidth
\epsfbox{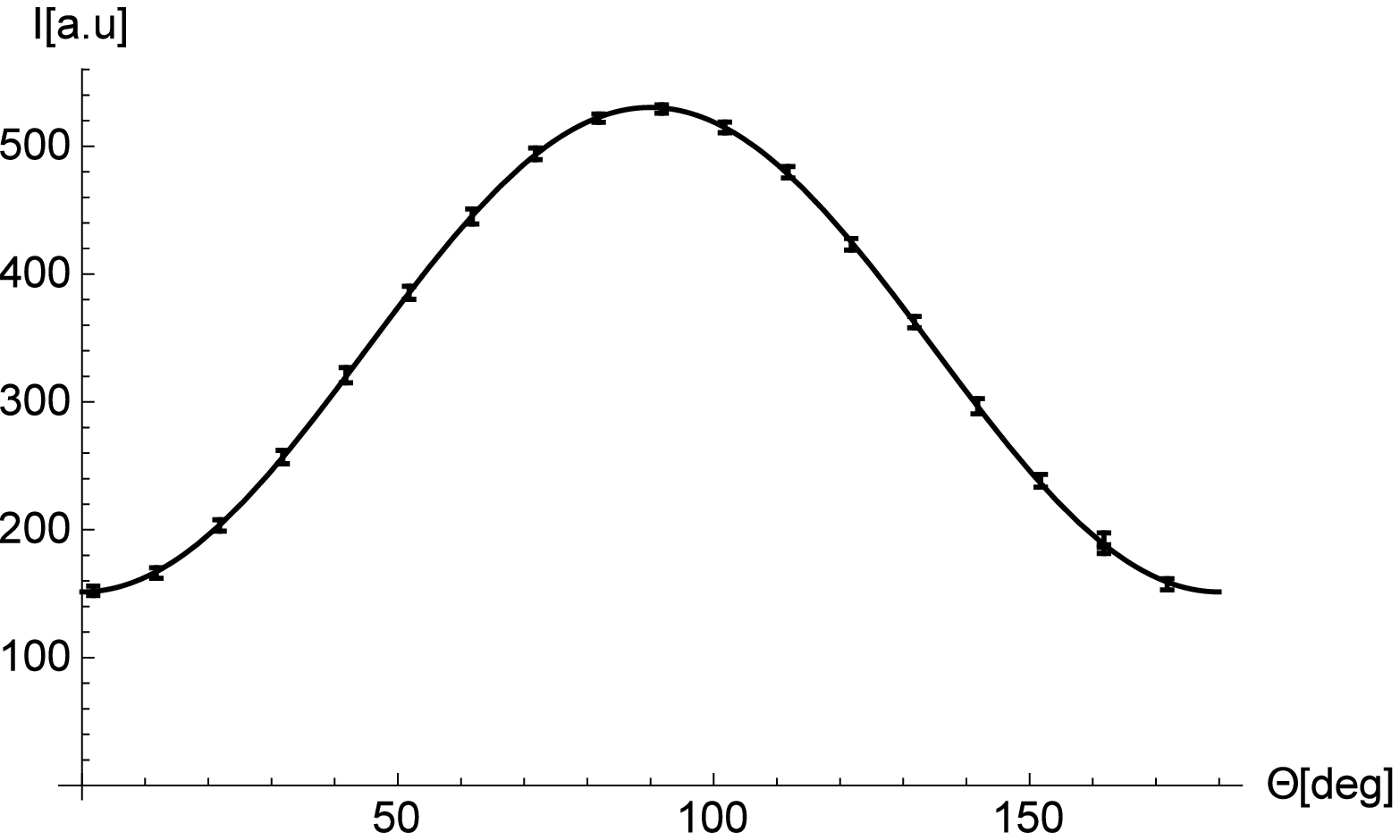}}
\vspace{1cm}
Figure 3


\begin{thebibliography}{99}
\bibitem{Jones} The expression of Jones vectors and Jones matrices depends on the convention of the complex amplitude. Throughout the paper the manner in Ref.~\cite{Hect} is employed.
\end{thebibliography}
\end{document}